\documentclass[a4paper,conference]{IEEEtran}
\pdfoutput=1
\IEEEoverridecommandlockouts
\usepackage{cite}
\usepackage{amsmath,amssymb,amsfonts}
\usepackage{algorithmic}
\usepackage{graphicx}
\usepackage{textcomp}
\usepackage{xcolor}
\usepackage{multirow}
\usepackage{makecell}
\def\BibTeX{{\rm B\kern-.05em{\sc i\kern-.025em b}\kern-.08em
    T\kern-.1667em\lower.7ex\hbox{E}\kern-.125emX}}
    
\begin{document}

\title{Parkinson's Disease Detection Using Ensemble Architecture from MR Images*\\
{\footnotesize \textsuperscript{*}}
\thanks{Supported by Multimedia Research Center (MRC), Department of Computing Science, University of
Alberta, Edmonton, Canada x.}
}

\author{\IEEEauthorblockN{1\textsuperscript{st} Tahjid Ashfaque Mostafa}
\IEEEauthorblockA{\textit{Department of Computing Science} \\
\textit{University of Alberta}\\
Edmonton, Canada \\
tahjid@ualberta.ca}
\and
\IEEEauthorblockN{2\textsuperscript{nd} Irene Cheng}
\IEEEauthorblockA{\textit{Department of Computing Science} \\
\textit{University of Alberta}\\
Edmonton, Canada \\
locheng@ualberta.ca}
\and
}

\maketitle
\begin{abstract}
Parkinson's Disease(PD) is one of the major nervous system disorders that affect people over 60. PD can cause cognitive impairments. In this work, we explore various approaches to identify Parkinson's using Magnetic Resonance (MR) T1 images of the brain. We experiment with ensemble architectures combining some winning Convolutional Neural Network models of ImageNet Large Scale Visual Recognition Challenge (ILSVRC) and propose two architectures. We find that detection accuracy increases drastically when we focus on the Gray Matter (GM) and White Matter (WM) regions from the MR images instead of using whole MR images. We achieved an average accuracy of 94.7\% using smoothed GM and WM extracts and one of our proposed architectures. We also perform occlusion analysis and determine which brain areas are relevant in the architecture decision making process.
\end{abstract}

\begin{IEEEkeywords}
Parkinson's Disease Detection, Ensemble Learning, Deep Learning, Magnetic Resonance Imaging
\end{IEEEkeywords}

\section{Introduction}
Parkinson's Disease(PD) is one of the most common neurodegenerative movement disorders with around 6.2 Million affected globally~\cite{sixpointtwomillion}. PD is mainly caused by loss of nerve cells (neurons) in Substantia Nigra, which is in the Basal Ganglia region of the brain. Neurons in this part of the brain are responsible for producing an organic chemical known as dopamine, which acts as a neurotransmitter in the brain by helping the neurons communicate. If the amount of dopamine produced in the brain is not sufficient, communication between neurons for coordinating body movement is hampered; causing PD, which has multiple neurological and motor symptoms like tremors, balance issues, Bradykinesia, impaired gait, stooped posture, speech problems etc. PD has been clinically defined and studied for decades, but it's exact mechanisms are still unclear~\cite{unclear_Cause}. PD is usually diagnosed with the manifestation of motor symptoms, which might not appear until a patient has lost 50-70\% of their neurons ~\cite{5070neurons}. Although a guaranteed cure for PD has not been found yet, early detection might play a crucial role in slowing or stopping the progression of the disease. Some new forms of treatment like Exenatide~\cite{exenatide}  show promising results in case of early detection. Analyzing the structural changes in the brain using Medical Imaging techniques have been proven to be helpful for detecting neurodegenerative diseases with cognitive impairments~\cite{medicalimaging1}~\cite{medicalimaging2}. In particular, Magnetic Resonance (MR) images provide better performance in brain structure analysis because of having high contrast and resolution within soft tissue. In this work, we propose two ensemble architectures to identify PD from MR images of the brain, one of which achieves comparable accuracy to existing state-of-the-art models. We also identify which region of the brain the network is focusing on using occlusion analysis.
\section{Related Works}
Over the years, various Machine Learning (ML)~\cite{unclear_Cause,relatedwork1,paz,relatedwork2} and Deep Learning (DL)~\cite{choi,relatedwork3} based approaches have been introduced for detecting Parkinson's Disease. Pazhanirajan et al. ~\cite{paz} used a Radial Basis Function Neural Network (RBFNN) for PD classification. Focke et al.~\cite{relatedwork1} used a SVM Classifier with Gray Matter (GM) and White Matter (WM) extracted from MR images. They reported 39.53\% and 41.86\% classification accuracy for GM and WM respectively.
Babu et al.~\cite{babu} proposed a Computer Aided Diagnosis (CAD) system. They achieved a 87.21\% accuracy in classifying PD using GM and identified Superior Temporal Gyrus as a potential biomarker. Choi et al.~\cite{choi} used Convolutional Neural Network[CNN] and SPECT Imaging to achieve an accuracy of 96\%. But SPECT Imaging requires injecting a radioactive tracer into the patient so it is invasive and not very popular. Over one year period in the NHS operation in England, around 100 times more MRI scans were performed compared to SPECT. Thus the approach seems to be impractical for normal medical use despite its reported high accuracy. Moreover the dataset suffers from class imbalance as about 69\% of their data is from PD patients. Class imbalance leads to models over classifying the majority class~\cite{classimbalance}.
Long et al.~\cite{long} used a ML based approach to detect PD from resting-state functional MRI (rsf-MRI). rsf-MRI detects subtle changes in blood flow whereas Structural MRI (sMRI) only captures the anatomical details and ignores all activity. Although they achieve 87\% accuracy, the dataset used by them was very small. Rana et al.~\cite{rana} used t-test feature selection on WM, GM and Cerebrospinal Fluid (CSF) and used a SVM for classification achieving 86.67\% accuracy for GM and WM and 83.33\% accuracy for CSF. In their other work~\cite{rana1} they considered the relation between tissues instead of considering the tissues separately and achieved an accuracy of 89.67\%. Braak’s neuroanatomical model of Parkinson’s Disease~\cite{braak} identifies the Substantia Nigra (SN) region as having significant correlation with PD and it is often used a Region Of Interest (ROI) in PD identification.

None of the methods we looked into were using ensemble architectures for PD detection and the detection accuracy using sMRI was lacking. We believed creating ensemble architectures combining different characteristics of deep learning models might give us better results. We describe our approach in the next section.

\section{Proposed Method}
\subsection{Data}
For our experiments, we used Parkinson Progression Markers Initiative (PPMI) dataset~\cite{ppmi}. It consists of T1-weighted sMRI scans for 568 PD and Healthy Control(HC) subjects from which we choose 445 subjects and discard the rest due to some structural anomalies during preprocessing steps. The resulting data has a class imbalance with 299 PD and 146 HC subjects. To address this issue, we collect 153 HC T1-weighted sMRI scans from the publicly available IXI dataset~\cite{ixi}. Our final dataset has is class balanced with 598 subjects. The demographic for the dataset is presented in Table~\ref{demotable}. 
\begin{table}[htbp]
\centering
\setlength{\tabcolsep}{5pt}
\renewcommand{\arraystretch}{1}
\caption{Demographic Data}\label{demotable}
\begin{tabular}{l|c|c|c}
 &  PD & HC & Average\\
\hline
Age(Years) &  $62.0 \pm 9.54$ & $49.2 \pm 16.9$ & $55.6 \pm 15.1$\\\hline
Sex (Male / Female) &  189 / 110 & 172 / 127 & 361 / 237\\
\end{tabular}
\end{table}

\subsection{Preprocessing}
There are morphological and dimensional differences between data since the scans come from different machines. To make the data comparable we had to standardize it to a common format, all scans were resized to the same dimensions. For preprocessing, Statistical Parameter Mapping (SPM12)~\cite{spm,spm1} and Computational AnatomTtoolbox (CAT12)~\cite{cat12} were used.

\begin{figure}[htbp]
\centering
\includegraphics[width=0.5\textwidth]{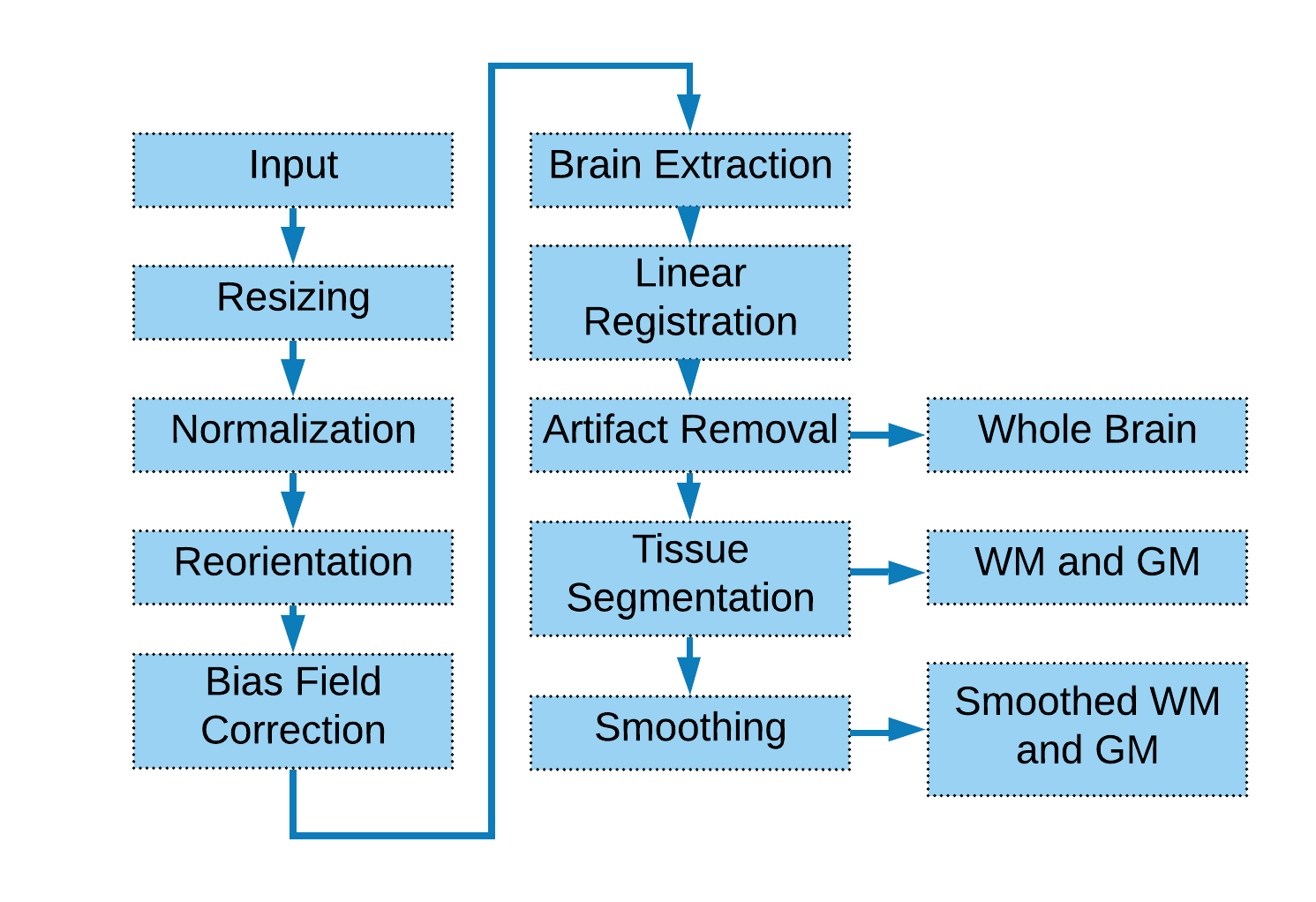}
\caption{Preprocessing Pipeline} \label{fig1}
\end{figure}

\begin{figure}[htbp]
    \centering
    \begin{tabular}{r}
        \includegraphics[width=0.4\textwidth]{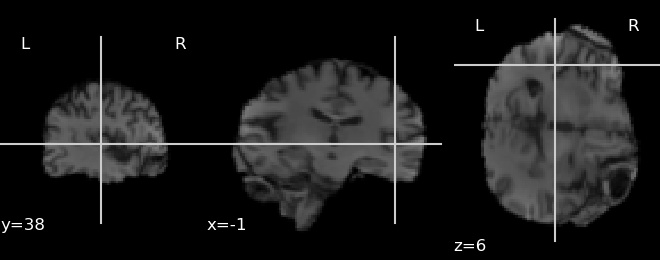}\\
        \multicolumn{1}{c}{\small (a) Whole Brain} \\
        \includegraphics[width=0.4\textwidth]{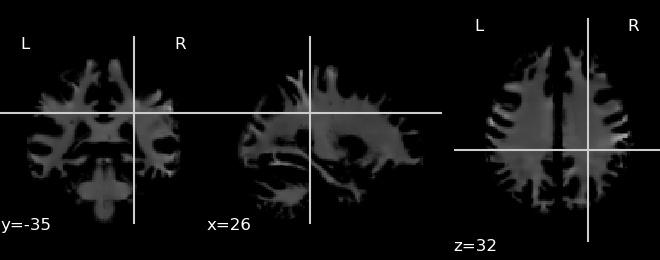}\\
        \multicolumn{1}{c}{\small (b) Extracted White Matter}\\
        \includegraphics[width=0.4\textwidth]{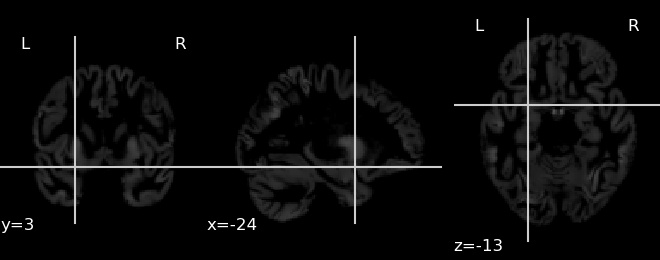}\\
        \multicolumn{1}{c}{\small (c) Extracted Gray Matter} {}   
    \end{tabular}
\caption{Sample MRI scans for a Healthy Control Patient and the extracted GM and WM}\label{fig2}

\end{figure}

Fig. ~\ref{fig1} shows the structure of our preprocessing pipeline. MRI intensity varies from subject to subject. To minimize discrepancies we normalize the values to [0,1]. Then all images were aligned to a standard space named Montreal Neurological Institute (MNI). Then a bias field correction (FAST)~\cite{biasfield} is performed to remove general intensity non-uniformities. FNIRT / BET~\cite{brainextract} was used to extract brain from the scans removing the skull, fat and background regions which do not contain useful information. The data was registered to MNI152 format (FLIRT)~\cite{registration,registration1}. After that artifact removal was performed, i.e. any voxel intensity values higher than 1 is corrected to be in the range [0,1]. Then a deformation method was applied to extract Gray Matter (GM) and White Matter (WM) from the scan and a 8mm Isotropic Gaussian Kernel was used to smooth and increase the signal-to-noise ratio and remove unnecessary portions of the scan. Finally we have three separate datasets: whole brain scans, GM and WM extracted from the brain and Smoothed GM and WM. An example of the extracted brain and the resultant WM and GM extracted from the brain is given in Fig. ~\ref{fig2}.

\subsection{Model}
We created two separate ensemble architectures combining existing models implemented in Pytorch \cite{pytorch}. We selected 6 models to construct our ensemble architecture:
\begin{itemize}
  \item ResNet 101
  \item SqueezeNet 1.1
  \item DenseNet 201
  \item VGG 19
  \item MobileNet V2
  \item ShuffleNet V2
\end{itemize}
These models are available from Torchvision \cite{torchvision} in two versions: without any kind of training (untrained) and trained on the ImageNet~\cite{imagenet} dataset. We used both untrained and pre-trained models to construct our ensemble networks and compared the performances of the resultant networks. The models are designed for the ImageNet\cite{imagenet} dataset, so we needed to modify them in order to accommodate our MRI data. The input layers of all models were changed to accommodate the shape of our input and the output layers were changed to predict between 2 classes (PD and HC) instead of the 1000 imagenet classes. 
\subsubsection{Ensemble model for Whole Brain Scan: Model 1}
\begin{figure*}[htbp]
\centering
  \includegraphics[width=.6\textwidth]{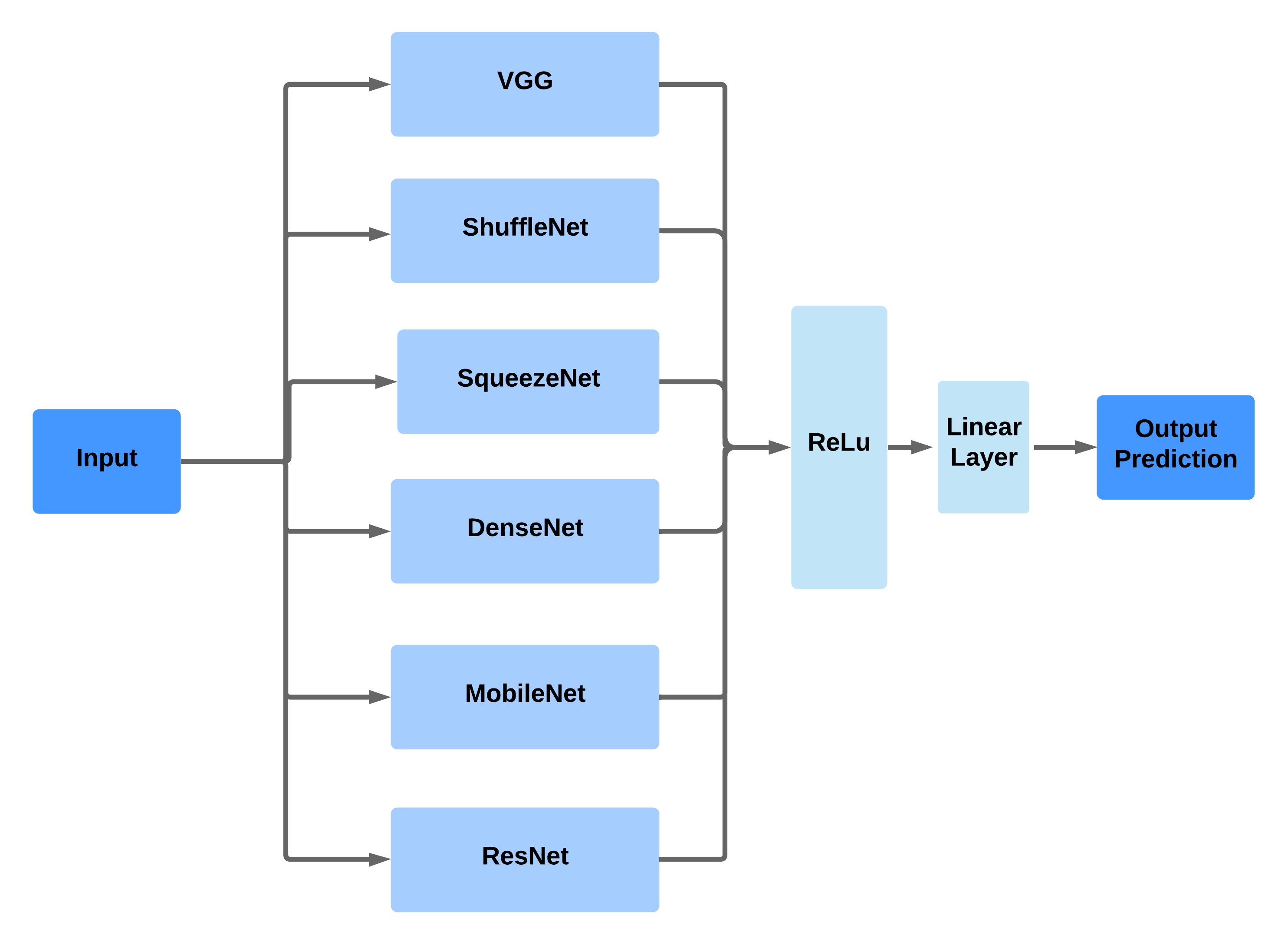}
  \medbreak
  \caption{Ensemble architecture for whole brain scan: Model 1}\label{fig3}
\end{figure*}

In this architecture, we pass our whole brain scan of dimensions $91 \times 109 \times 91$ through six models in parallel and the concatenated output is passed through a Rectified Linear Unit (ReLU) activation function and then to a Linear layer with 2 output classes. Fig. ~\ref{fig3} shows a visual representation of this architecture.
\subsubsection{Ensemble model for Extracted GM and WM Scans: Model 2}
\begin{figure*}[htbp]
\centering
  \includegraphics[width=.7\textwidth]{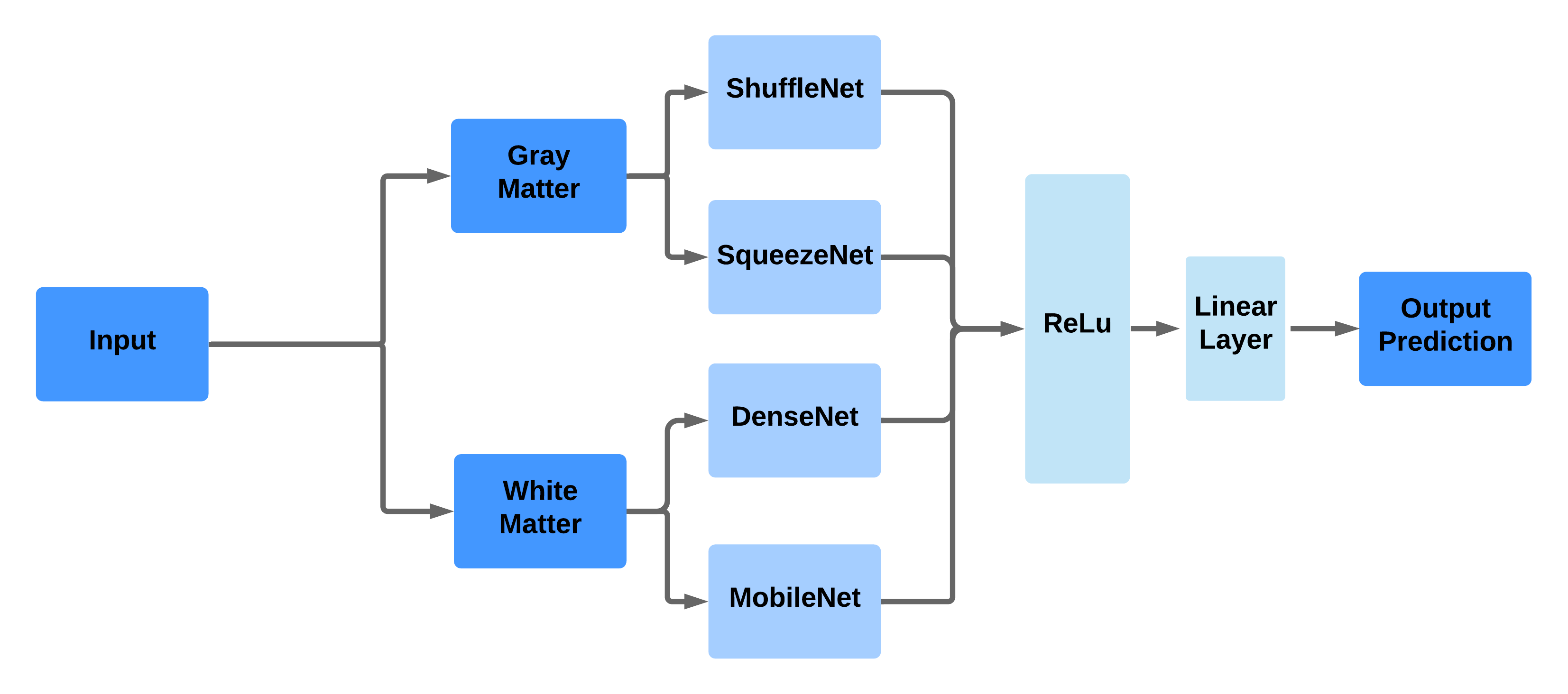}
  \medbreak
  \caption{Ensemble architecture for Extracted GM and WM Scans : Model 2}\label{fig4}
\end{figure*}

This architecture is trained on the extracted GM and WM scans of dimension $121 \times 145 \times 121$. It is comprised of four models. The GM scans are passed through ShuffleNet and SqueezeNet and the WM scans are passed through DenseNet and MobileNet. The output of all models are concatenated and passed through a ReLU activation layer and a linear layer with 2 output classes to get final predictions.
Fig. ~\ref{fig4} shows a visual representation of this architecture.

\section{Experimental Results}
Two versions were constructed for each of our ensemble architectures; one with all untrained constituent models and another with all pretrained constituent models. The dataset was split randomly, and 80\% was selected for training and 20\% for testing. Each model was trained for 25 epochs with an Adam Optimizer and Cross Entropy Loss function. At each epoch, the training set was further split randomly, and 20\% was selected for validation. We repeated the procedures listed above 5 times to obtain an average for 2 different learning rates \textbf{.001} and \textbf{.0001}. The results are presented in Table~\ref{resultstable}. 
\begin{table*}[htbp]
\centering
\setlength{\textwidth}{15pt}
\renewcommand{\arraystretch}{2}
\caption{Results}\label{resultstable}
\begin{tabular}{@{\extracolsep{\fill}} c | c | c | c | c }

 \thead{Model} & \thead{Use \\Smoothed \\ Scan} & \thead{Pre Trained} & \thead{Learning Rate} & \thead{Classification \\Accuracy\\(On a scale of 0-1)}\\
\hline
 \multirow{4}{*}{\makecell{Model 1 \\ Whole Brain Scans}}&  \multirow{4}{*}{N/A} &\multirow{2}{*}{False} & .001 & $\textbf{0.7617} \pm \textbf{0.0041}$\\\cline{4-5}

 &   & &.0001 & $0.7459 \pm 0.0042$\\\cline{3-5}

 &   &\multirow{2}{*}{True} &.001 & $0.6800 \pm 0.0113$\\\cline{4-5}
 &   & &.0001 & $0.6613 \pm 0.0311$\\\hline
 \multirow{8}{*}{\makecell{Model 2 \\ Extracted GM and WM \\Scans}}&  \multirow{4}{*}{False} &\multirow{2}{*}{False} & .001 & $0.5487 \pm 0.0002$\\\cline{4-5}

 &   & &.0001 & $0.6847 \pm 0.0093$\\\cline{3-5}

 &   &\multirow{2}{*}{True} &.001 & $0.9231 \pm 0.0258$\\\cline{4-5}
 &   & &.0001 & $\textbf{0.9366} \pm \textbf{0.0170}$\\
  \cline{2-5} &  \multirow{4}{*}{True} &\multirow{2}{*}{False} & .001 & $0.5410 \pm 0.0106$\\\cline{4-5}

 &   & &.0001 & $0.7276 \pm 0.0476$\\\cline{3-5}

 &   &\multirow{2}{*}{True} &.001 & $	0.9291 \pm 0.0170$\\\cline{4-5}
 &   & &.0001 & $\textbf{0.9470} \pm \textbf{0.0083}$\\\hline
 Focke et al.\cite{relatedwork1} [GM]&N/A&N/A&N/A&0.3953\\\hline
  Focke et al.\cite{relatedwork1} [WM]&N/A&N/A&N/A&0.4186\\\hline
   Babu et al.\cite{babu} [GM]&N/A&N/A&N/A&0.8721\\\hline
    Rana et al.\cite{rana} [GM \& WM]&N/A&N/A&N/A&0.8667\\\hline
        Rana et al.\cite{rana1}&N/A&N/A&N/A&0.8967\\\hline
\end{tabular}
\end{table*}

For whole brain scans, the best results we achieved was an average accuracy of $0.7617 \pm 0.0041$ with a Learning Rate of .001. As shown in Table~\ref{resultstable}, using pre trained models as constituents of our architecture does not help our predictions in this case. 

We can see a drastic increase in performance when we use the extracted GM and WM scans with pre-trained constituent models for Model 2. The best average accuracy for non-smoothed regions is $0.9366 \pm 0.0170 $. 

For the same model architecture, we get even better average accuracy of $0.9470\pm0.0083$ by using Smoothed GM and WM extractions, which significantly outperforms all the existing work we mentioned on sMRI for PD identification. Since this is the best accuracy we have achieved, we used this method to perform occlusion analysis and identify the Region Of Interest (ROI) according to the method proposed by Rieke et al.~\cite{visualize}. 
\subsubsection{Occlusion Analysis}
\begin{figure*}[htbp]
    \centering
    \begin{tabular}{r}
        \includegraphics[width=\textwidth]{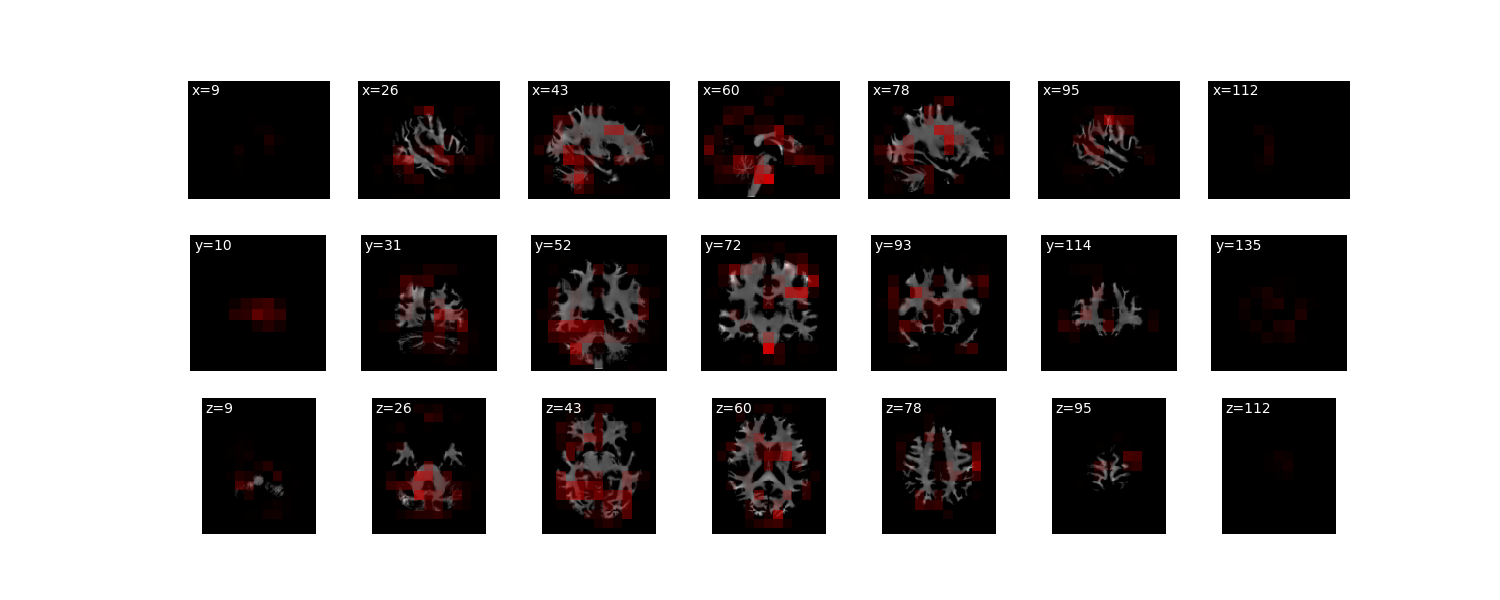} \\
        \multicolumn{1}{c}{\small (a) Heatmap for White Matter} \\
        \includegraphics[width=\textwidth]{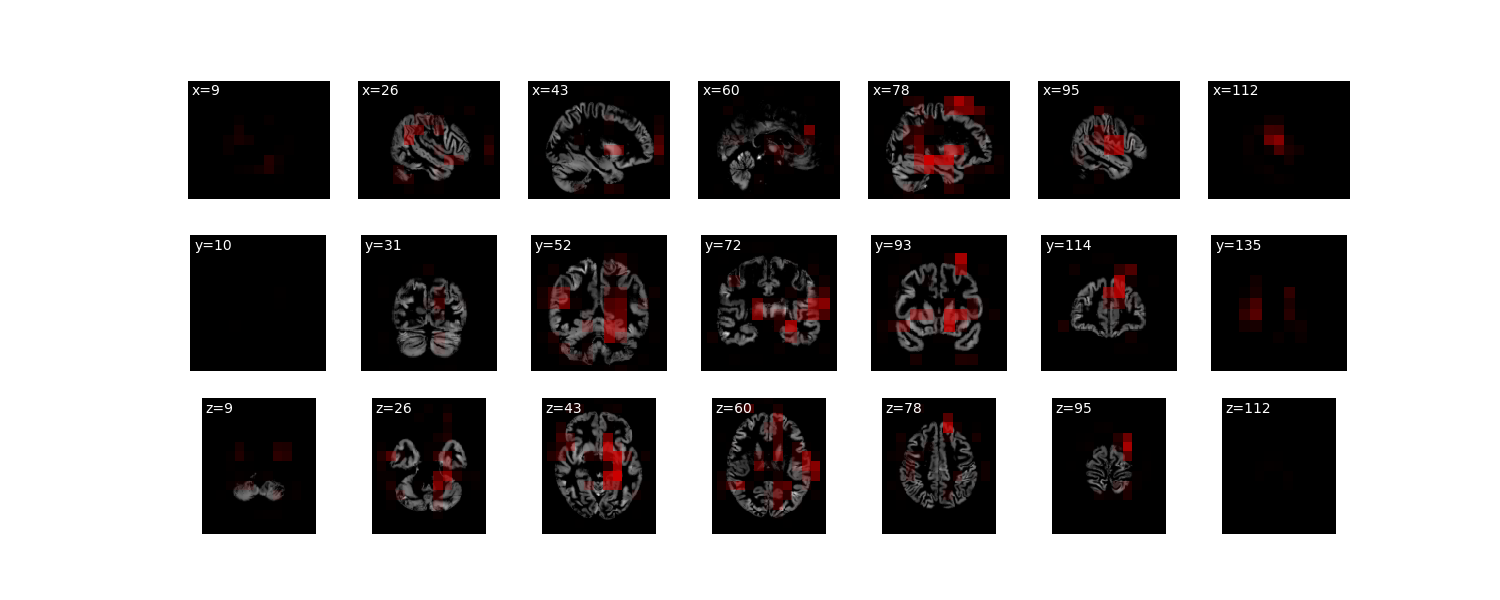} \\
        \multicolumn{1}{c}{\small (b) Heatmap for Gray Matter} \\
    \end{tabular}
\caption{Relevance Heatmaps for Occlusion of GM and WM images}\label{heatmap}
\end{figure*}

To understand which regions of the brain are important in the decision making process we perform a slightly modified version of occlusion analysis proposed in ~\cite{visualize} to fit our data. In this analysis, usually a part of the scan is occluded with gray or white patch and the output from the network is recalculated. The occluded region is considered to be important if the probability of the target class decreases compared to the original image. The heatmap of relevance is calculated by sliding the patch across the image and plotting the difference in the probability. 
In our case we repeated the procedure twice, once for each of the GM and WM images. The resultant heatmaps are presented in Fig. ~\ref{heatmap}, where the important areas are marked in red.

% \begin{figure*}[t]
%     \centering
%     \begin{subfigure}[b]{\textwidth}
%         \centering
%         \includegraphics[height=2in]{figure_wm.png}        
%         \caption{a.) Heatmap for White Matter}
%     \end{subfigure}%
%     ~ 
    
%     \begin{subfigure}[b]{\textwidth}
%         \centering
%         \includegraphics[height=2in]{figure_gm.png}
%         \caption{b.) Heatmap for Gray Matter}
%     \end{subfigure}

% \caption{Relevance Heatmaps for Occlusion of GM and WM images}\label{heatmap}

\begin{figure*}[htbp]
    \centering
        \includegraphics[width=.8\textwidth]{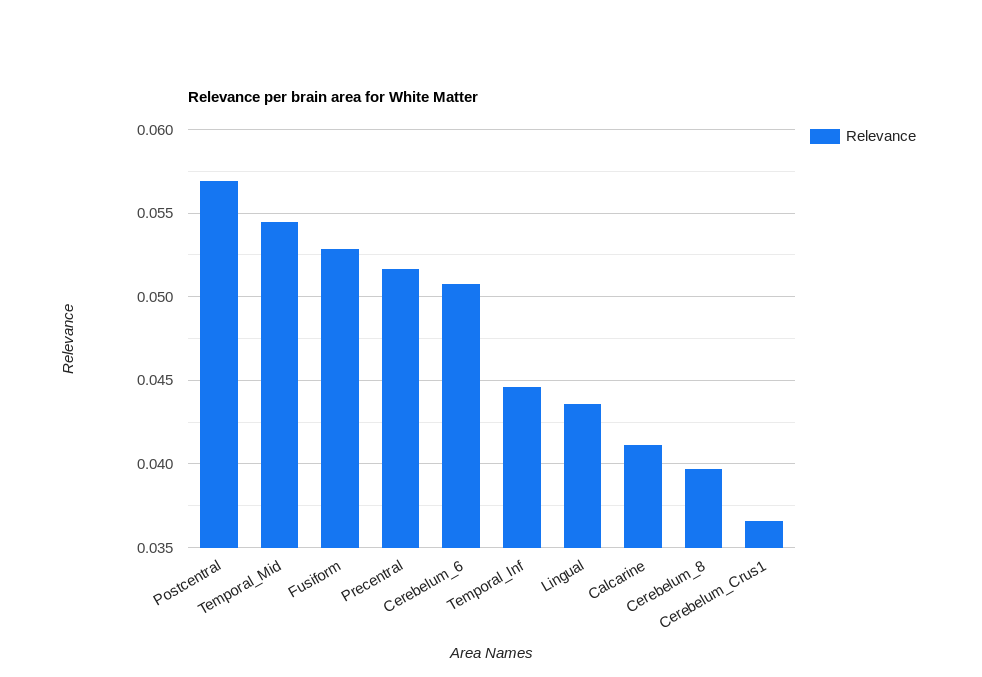} 
        \caption{Relevance per brain area for White Matter}\label{relevancewm}
        \includegraphics[width=0.8\textwidth]{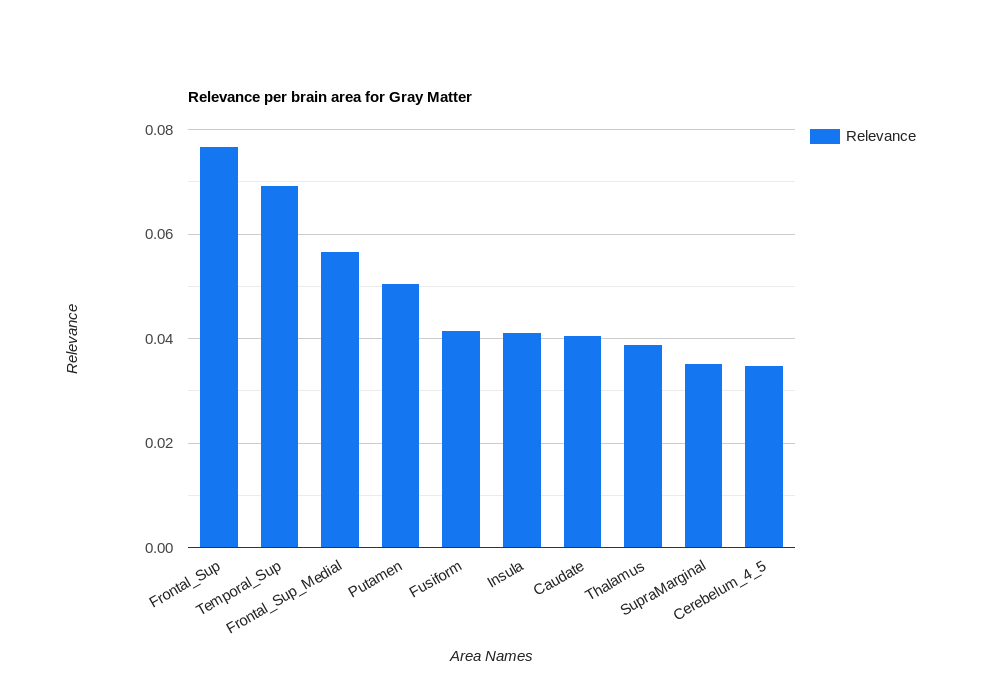}
        % \multicolumn{1}{c}{\small (a) White Matter} &
        % \multicolumn{1}{c}{\small (b) Gray Matter} \\
\caption{Relevance per brain area for Gray Matter}\label{relevancegm}
\end{figure*}

% \end{figure*}
To better understand the above heatmaps we also calculated the relevance per brain area using methods provided by~\cite{visualize} which is presented in Fig. ~\ref{relevancewm} and Fig. ~\ref{relevancegm}. We can see for Gray Matter the most focused on area is Superior Frontal Gyrus and for White Matter it is Postcentral Gyrus.
\section{Conclusion and Future Works}
In this paper we proposed two novel deep learning architectures for Parkinson's Disease Detection using Ensemble Learning, which outperforms related works. We achieved higher accuracy using smoothed GM and WM with our second ensemble architecture. We also applied occlusion analysis to identify the region of significance for model decision. Our next step will be to analyse the decision making process of our model using the Substantia Nigra as the target of significance to assess if it produces better results than using GM and WM regions

% \begin{figure*}[t!]
%     \centering
%     \begin{subfigure}[t]{.5\textwidth}
%         \centering
%         \includegraphics[height=1.5in]{bar-graph-WM.png}        
%         \caption{a.) White Matter}
%     \end{subfigure}%
%     ~ 
%     \begin{subfigure}[t]{.5\textwidth}
%         \centering
%         \includegraphics[height=1.5in]{bar-graph-GM.png}
%         \caption{b.) Gray Matter}
%     \end{subfigure}

% \caption{Relevance per brain area of GM and WM images}\label{relevance}

% \end{figure*}
\section*{Acknowledgment}
Special thanks to PPMI for supporting Parkinson’s disease research by maintaining and updating their clinical dataset.

Data used in the preparation of this article were obtained from the Parkinsons
Progression Markers Initiative (PPMI) database (www.ppmi-info.org/data). For
up-to-date information on the study, visit www.ppmi-info.org.

PPMI a public-private partnership is funded by the Michael J. Fox Founda-
tion for Parkinsons Research and funding partners found at www.ppmi-info.org/fundingpartners.

Special thanks for the guidance from Dr. Sara Soltaninejad (soltanin@ualberta.ca), previous PhD student at the Department of Computing Science, University of Alberta.

Financial support from the Natural Sciences and Engineering Research Council of Canada (NSERC) is gratefully acknowledged. 

% \begin{figure}[H]
%         \includegraphics[width=0.2\textwidth]{nserc.png} 
% \end{figure}

\vspace{12pt}
\end{document}